\documentclass[12pt]{article}

\usepackage{mathrsfs}
\usepackage[T1]{fontenc}
\usepackage{mathpazo}
\usepackage{setspace}
\usepackage{amsfonts}
\usepackage{amssymb}
\usepackage{amsmath}
\usepackage{epsfig}
\usepackage{latexsym}
\usepackage{color}
\usepackage{graphicx}
\usepackage{nicefrac}
\usepackage[latin1]{inputenc}
\usepackage{pstricks}
\usepackage{slashed}
\usepackage{multirow}

\usepackage{hyperref}

\def\hybrid{\topmargin 0pt \oddsidemargin 0pt 
        \headheight 0pt \headsep 0pt
        \textwidth 16,5cm 
        \textheight 23cm 
        \marginparwidth .875in
        \parskip 5pt plus 1pt \jot = 1.5ex}

\hybrid

\catcode`\@=11

\def\marginnote#1{}
\newcount\hour
\newcount\minute
\newtoks\amorpm
\hour=\time\divide\hour by60
\minute=\time{\multiply\hour by60 \global\advance\minute by-\hour}
\edef\standardtime{{\ifnum\hour<12 \global\amorpm={am}%
        \else\global\amorpm={pm}\advance\hour by-12 \fi
        \ifnum\hour=0 \hour=12 \fi
        \number\hour:\ifnum\minute<10 0\fi\number\minute\the\amorpm}}
\edef\militarytime{\number\hour:\ifnum\minute<10 0\fi\number\minute}

\def\draftlabel#1{{\@bsphack\if@filesw {\let\thepage\relax
   \xdef\@gtempa{\write\@auxout{\string
      \newlabel{#1}{{\@currentlabel}{\thepage}}}}}\@gtempa
   \if@nobreak \ifvmode\nobreak\fi\fi\fi\@esphack}
        \gdef\@eqnlabel{#1}}
\def\@eqnlabel{}
\def\@vacuum{}
\def\draftmarginnote#1{\marginpar{\raggedright\scriptsize\tt#1}}

\def\draft{\oddsidemargin -.5truein
        \def\@oddfoot{\sl preliminary draft \hfil
        \rm\thepage\hfil\sl\today\quad\militarytime}
        \let\@evenfoot\@oddfoot \overfullrule 3pt
        \let\label=\draftlabel
        \let\marginnote=\draftmarginnote
   \def\@eqnnum{(\theequation)\rlap{\kern\marginparsep\tt\@eqnlabel}%
\global\let\@eqnlabel\@vacuum} }

\def\draft2{
        \def\@oddfoot{\sl preliminary draft \hfil
        \rm\thepage\hfil\sl\today\quad\militarytime}
        \let\@evenfoot\@oddfoot \overfullrule 3pt
        \let\label=\draftlabel
        \let\marginnote=\draftmarginnote
   \def\@eqnnum{(\theequation)\rlap{\kern\marginparsep\tt\@eqnlabel}%
\global\let\@eqnlabel\@vacuum} }

\def\preprint{\twocolumn\sloppy\flushbottom\parindent 2em
        \leftmargini 2em\leftmarginv .5em\leftmarginvi .5em
        \oddsidemargin -.5in \evensidemargin -.5in
        \columnsep .4in \footheight 0pt
        \textwidth 10.in \topmargin -.4in
        \headheight 12pt \topskip .4in
        \textheight 6.9in \footskip 0pt
        \def\@oddhead{\thepage\hfil\addtocounter{page}{1}\thepage}
        \let\@evenhead\@oddhead \def\@oddfoot{} \def\@evenfoot{} }

\def\numberbysection{\@addtoreset{equation}{section}
        \def\theequation{\thesection.\arabic{equation}}}

\def\underline#1{\relax\ifmmode\@@underline#1\else
        $\@@underline{\hbox{#1}}$\relax\fi}

\def\titlepage{\@restonecolfalse\if@twocolumn\@restonecoltrue\onecolumn
     \else \newpage \fi \thispagestyle{empty}\c@page\z@
        \def\thefootnote{\fnsymbol{footnote}} }

\def\endtitlepage{\if@restonecol\twocolumn \else \newpage \fi
        \def\thefootnote{\arabic{footnote}}
        \setcounter{footnote}{0}} 

\catcode`@=12
\relax

\def\figcap{\section*{Figure Captions\markboth
        {FIGURECAPTIONS}{FIGURECAPTIONS}}\list
        {Figure \arabic{enumi}:\hfill}{\settowidth\labelwidth{Figure
999:}
        \leftmargin\labelwidth
        \advance\leftmargin\labelsep\usecounter{enumi}}}
 \relax
\def\tablecap{\section*{Table Captions\markboth
        {TABLECAPTIONS}{TABLECAPTIONS}}\list
        {Table \arabic{enumi}:\hfill}{\settowidth\labelwidth{Table
999:}
        \leftmargin\labelwidth
        \advance\leftmargin\labelsep\usecounter{enumi}}}
 \relax
\def\reflist{\section*{References\markboth
        {REFLIST}{REFLIST}}\list
        {[\arabic{enumi}]\hfill}{\settowidth\labelwidth{[999]}
        \leftmargin\labelwidth
        \advance\leftmargin\labelsep\usecounter{enumi}}}
 \relax
\makeatletter
\newcounter{pubctr}
\def\publist{\@ifnextchar[{\@publist}{\@@publist}}
\def\@publist[#1]{\list
        {[\arabic{pubctr}]\hfill}{\settowidth\labelwidth{[999]}
        \leftmargin\labelwidth
        \advance\leftmargin\labelsep
        \@nmbrlisttrue\def\@listctr{pubctr}
        \setcounter{pubctr}{#1}\addtocounter{pubctr}{-1}}}
\def\@@publist{\list
        {[\arabic{pubctr}]\hfill}{\settowidth\labelwidth{[999]}
        \leftmargin\labelwidth
        \advance\leftmargin\labelsep
        \@nmbrlisttrue\def\@listctr{pubctr}}}
 \relax
\makeatother

\def\ba{\begin{equation}}
\def\ea{\end{equation}}

\def\k{\kappa}

\def\a{\alpha}

\def\b{\beta}

\def\G{\Gamma}
\def\d{\delta}

\def\e{\epsilon}

\def\m{\mu}

\def\Om{\Omega}

\def\L{\Lambda}
\def\s{\sigma}

\def\cL{{\cal L}}

\def\no{\noindent}

\def\qq{\qquad}

\def\IR{\relax{\rm I\kern-.18em R}}

\def \ha {{1\over 2}}

\def \ov {\over}

\def\diag{{\rm diag}}

\begin{document}


\renewcommand{\theequation}{\thesection.\arabic{equation}}
\csname @addtoreset\endcsname{equation}{section}

\newcommand{\eqn}[1]{(\ref{#1})}
\newcommand{\be}{\begin{eqnarray}}
\newcommand{\ee}{\end{eqnarray}}
\newcommand{\non}{\nonumber}
\begin{titlepage}
\strut\hfill
\vskip 1.3cm
\begin{center}

\vskip -2 cm


\vskip 2 cm

{\large \bf Recent developments in non-Abelian T-duality in string theory}
\footnote{{\tt Proceedings contribution to the {\it 10th
Hellenic School on Elementary Particle Physics and Gravity}, Corfu,
Greece, September 2010. \hfill}}

\vskip 0.5in

{\bf Konstadinos Sfetsos}

\vskip 0.1in

Department of Engineering Sciences, University of Patras,\\
26110 Patras, Greece\\
{\footnotesize sfetsos@upatras.gr}

\vskip .1in

\vskip .15in

\end{center}

\vskip .4in

\centerline{\bf Synopsis}

\no
 We briefly review the essential points of our recent work in
non-Abelian T-duality. In particular, we show how non-abelian
T-duals can effectively describe infinitely high spin sectors of a
parent theory and how to implement the transformation in the
presence of non-vanishing Ramond fields in type-II supergravity.

\no


\end{titlepage}
\vfill
\eject



\def\baselinestretch{1.2}
\baselineskip 20 pt \no

\section{Prolegomena}

Abelian T-duality was originally formulated in a path integral
approach in which central r\^ole played the isometry group $G=U(1)$
of the background one chooses to dualize
\cite{Buscher:1987sk}. When this group is
non-Abelian one may follow a similar path to naturally
arrive at the notion of non-Abelian T-duality
\cite{delaossa:1992vc}. However, the similarities between the two
cases stop here. In particular:

 \noindent
${\bf 1.}$ Unlike the Abelian case, the
non-Abelian T-duality transformation is non-invertible in the standard path integral
formulation since the isometries are no longer present in the T-dual
background (however, see comments at the end of this note).

\noindent
${\bf 2.}$ For compact commuting isometries one may argue that T-duality is actually
a true symmetry of string theory. There is no analogous statement for the non-Abelian cases.

\noindent
${\bf 3.}$ Even for compact groups,
the variables of the T-dual background are generically non-compact.

\noindent
${\bf 4.}$ The formulation of T-duality in the presence of Ramond fluxes presents technical difficulties.
In the Abelian case the unique dimensional reduction to nine dimensions
of the type-II supergravities
provided for the transformation rules \cite{Bergshoeff:1995as}.
This possibility hasn't been explored for non-Abelian T-duality.

\noindent
In this note we summarize the essential points of recent developments in the subject.

\section{Pure NS backgrounds}

This section is based mainly on \cite{Polychronakos:2010hd} and on general techniques developed in
\cite{Polychronakos:2010fg}.
Consider a pure NS background
described by a metric $G_{\mu\nu}$, an antisymmetric tensor $B_{\mu\nu}$ and
a dilaton $\Phi$.
These can couple in a classical two-dimensional $\s$-model action for the target space variables $X^\m$.
Let's denote this action by $S(X)$ and the corresponding theory by $\cal C$. We will also
assume that there is an isometry group $G$ leaving the action invariant with the variables
$X^\m$ transforming accordingly.
We gauge a subgroup $H\subset G$ by introducing,
in a Buscher-like approach,
gauge fields $A_\pm \in \cL(H)$
and a Lagrange multiplier term for their field strength $F_\pm$. The corresponding action is
\be
S_{\rm T\!-\!dual}(X,v,A_\pm)  = S_{\rm g}(X,A_\pm) - i\int d^2\s {\rm Tr}(v F_{+-})\ ,\qquad
S_{\rm g}(X,0)=S(X)\ .
\label{stdd}
\ee
This action should be
invariant under
$
A_\pm \to \L^{-1} (A_\pm - \partial_\pm) \L$ and $ v \to \L^{-1} v \L
$,
together with the transformation of the $X^\m$'s.
The gauge fields enter at most quadratically and non-dynamically in $S_{\rm g}(X,A_\pm)$.
Integrating them out gives the T-dual $\s$-model
with transformed background fields.
We should also gauge fix $\dim(H)$ among the parameters $X^\m$ and the Lagrange multipliers in $v$.
The maximum number of entries in $v$ that can be gauge fixed cannot exceed $\dim(H)-{\rm rank}(H)$.

\noindent
A non-Abelian T-dual ia a generically non-compact manifold
even for compact isometries.
The corresponding $\s$-model
is the result of a delicate limit taken in
some parent theory. It effectively describes sector(s) of infinitely high quantum numbers
with a simultaneous stretching of coordinates in the corresponding parent background.
To see that consider adding to $S_{\rm g}(X,A_\pm)$, not the Lagrange multiplier term, but
the gauged WZW action $I(h,A_\pm)$ at level $\ell$
for the group element $h\in H$.
Expanding infinitesimally as
$
h = \mathbb{I} + i {v/ \ell} + \cdots
$
and taking the limit $\ell\to \infty$ one finds
that $I(h,A_\pm)$ precisely reproduces the
Lagrange multiplier term.
Hence, one may think of the non-Abelian dual of the original theory $\cal C$
with action $S(X)$ as the limit of the gauged
tensor product theory ${\cal C}\times H_\ell$, when $\ell\to \infty$.
The advantage of this point of view is that the original theory may be better suited to study
before the limit is taken. Also, the non-compactness of the dual variables in $v$ is naturally explained.

\noindent
The above classical statement can be promoted at the level of the quantum states of the theory.
Consider as the simplest example the non-Abelian T-dual of the
$SU(2)$ WZW model with respect to $SU(2)$ acting vectorially.
In this case $\cal C$
is a current algebra theory and $S(X)$ is the associated WZW model action.
The T-dual background has zero antisymmetric tensor. The
metric and dilaton read
\be
&& ds^2  =  d\psi^2 + {\cos^2 \psi\ov x_3^2} dx_1^2
+ {\left( x_3 dx_3 + (\sin\psi\cos\psi +  x_1+\psi) dx_1\right)^2\ov x_3^2\cos^2\psi}\ ,
\nonumber\\
&& \Phi  = - \ln (x_3 \cos\psi)\ ,
\label{ghd1}
\ee
where $\psi$ is periodic and $x_1, x_3$ are non-compact,
a background possessing no isometries.
It describes the infinitely large spin sector
of the $SU(2)_{k_1}\times SU(2)_{k_2}/SU(2)_{k_1+k_2}$ coset CFT model.
In Physics it is important to be able to solve the field equations in a given background
especially the scalar wave equation.
Doing that by traditional methods is hopeless given the complexity of the background.
Making use of the underlying CFT, the general state can be written
as a multiple sum
involving the Clebsch--Gordan coefficients for a state $|j,m\rangle $
in the diagonal $SU(2)$ composed from states $|j_1,m_1\rangle |j_2,m_2\rangle$
in $SU(2)\times SU(2)$ for the
left and the right sectors separately and the Wigner's $d$-functions,
is such a way that a singlet of the diagonal
$SU(2)_L\times SU(2)_R$ is formed.
The eigenvalues are (for $k_{1,2}\gg 1$)
\be
E^j_{j_1,j_2} = {j_1(j_1+1)\ov k_1} + {j_2(j_2+1)\ov k_2} - {j(j+1)\ov k_1+k_2}\ .
\ee
To illustrate how the high spin limit is taken assume that one of the spins is extremely large, i.e.
$j_1, j\gg 1$ and $j_2=$finite.
In this limit, the eigenvalues become infinite unless the level $k_1$ becomes large as well
and proportional to $j$. Specifically, let
$j_1= j-n$ and $k_1 = {k_2 j/ \d} $,
where $|n|\leqslant j_2$ and $\d\in \mathbb{R}^+$. Then
\be
E_{j_2,n,\d}= \lim_{j\to
\infty} E^j_{j_1,j_2} = {j_2(j_2+1)\ov k_2} +  {\d - 2 n\ov k_2}\ \d\ .
\label{iginfi}
\ee
It turns out that in the $k_1\to \infty$ limit the background of the coset model becomes
that in \eqn{ghd1}.
Also, the solutions of the scalar wave equation can be obtain
from a delicate limit of the corresponding
solutions of the coset model. For example, the states with $j_2=1/2$ are
\be
\Psi_{1/2,\pm 1/2,\d} = \pm { \b_3\ov \d v_3}\ \cos 2 \d v_3 + {2\d \b_0 v_3\mp \b_3 \ov 2 \d^2 v_3^2}\
\sin 2 \d v_3\ ,
\ee
where $v_3,\b_0$ and $\b_3$ are functions of the variables $x_1,x_2$ and $\psi$.
We couldn't have constructed this solution, let alone one for general spin $j$,
by directly solving the scalar wave equation for the background \eqn{ghd1}.

\section{Non-trivial RR backgrounds }

This section is based on \cite{Sfetsos:2010uq,Lozano:2011kb}.
In type-II supergravity it is necessary to know, in addition to the NS fields,
how Ramond fluxes transform under T-duality.
The left and right world sheet derivatives transform differently under
T-duality and this defines two orthonormal frames related by a Lorentz transformation matrix $\L$.
The induced action
on spinors is given by a matrix $\Omega$ obtained by
\be
\Omega^{-1}  \Gamma^i  \Omega =  \Lambda^i{}_j \Gamma^j\ .
\label{spino1}
\ee
The RR-fields are combined into
a bi-spinor according to which type-II supergravity they belong to as
\be
 {\rm IIB}:  \quad P = {e^{\Phi}\ov 2} \sum_{n=0}^4{\slashed{F}_{2n+1}\ov (2n+1)!} \ ,
\qquad ({\rm massive)\ IIA}:  \quad  P ={ e^{\Phi}\ov 2}
\sum_{n=0}^5 {\slashed{F}_{2n}\ov (2n)!}  \ ,
\ee
with
$
{\slashed F}_p =  \G^{\m_1\cdots \m_p} F_{\m_1\cdots \m_p}
$.
and where we have used the democratic formulation
of type-II supergravities where all forms up to order ten appear.
The fluxes transform under T-duality according to
\be
\hat P = P \Om^{-1}\ ,
\label{ppom}
\ee
where we have denoted by a hat the bi-spinor obtained after the duality.
The details of the matrix $\Omega$ depend on the case of interest.
For comparison, for Abelian T-duality
this is simply given by $\Om = \G_{11} \G_1$ \cite{Hassan:1999bv}, where $1$ labels
the isometry direction and $\G_{11}$ the product
of all Gamma matrices. In the Abelian case we flip between type-IIA and type-IIB,
but in non-Abelian cases we might change or stay within the same theory.

\noindent
Many interesting supergravity backgrounds
have as an essential part group or coset manifolds. Hence,
we concentrate on Principle Chiral-type models which can cover both cases as we will see.
Consider an group element $g\in G$ and the components
of the left invariant Maurer--Cartan forms $L^a_\m =-i\ {\rm
Tr}(t^a g^{-1}\partial_\m g)$. The representation matrices
$t^a$ obey the Lie algebra with structure constants
$f^{ab}{}_c$. The most general $\s$-model invariant under
the global symmetry $g\to g_0 g$, with $g_0\in G$, is (we ignore spectator fields)
\be
S = \frac{1}{2} \int d^2 \s  \ E_{ab} L^a_+ L^b_-\ ,\qquad L^a_\pm = L^a_\mu\partial_\pm X^\m\ .
\label{dualmods}
\ee
Consider first the case in which $a$ and $b$ run over the whole group.
Then $E$ is a $\dim(G)$ square invertible constant matrix.
It turns out that the T-dual $\s$-model with respect to the full $G$ symmetry group is
\be
\tilde{S} =  \frac{1}{2} \int d^2 \s \,  (M^{-1})^{ab} \partial_+ v_a \partial_- v_b\ ,
\quad M_{ab}  = E_{ab} +  f_{ab}\ ,\qquad f_{ab}=f_{ab}{}^c v_c\
\label{dualmodss}
\ee
and the induced dilaton is $\Phi = -\ha \ln \det M$.
The variables of the T-dual model are the Lagrange multipliers $v_a$ since it is possible
to gauged fix the group element $g$ to unity. This is possible since the left sided
group action acts with no isotropy.
The world-sheet derivatives transform as
\be
L_+^a = (M^{-1})^{ba} \partial_+ v_b\ ,\qquad L_-^a = -(M^{-1})^{ab}\partial_- v_b\ .
\ee
Denoting by $\eta=\k^T \k$ the symmetric part of $E$,
the frame relating Lorentz transformation is
\be
\L = -\k M^{-1T} M\k^{-1}\quad  \Longrightarrow \quad
\Om = e^{\ha \tilde f_{ab} \G^{ab}}
\prod_{i=1}^{\dim(G)}\! (\G_{11} \G_i) \ ,\quad \tilde f= \k^{-1T} (S+f) \k^{-1}\ .
\label{dhfi}
\ee
Clearly if the dimensionality of the duality group is even
then we stay in the same type-II supergravity theory, otherwise we flip from (massive) type--IIA
supergravity to type--IIB and vise versa.

\noindent
To extend the discussion for coset $G/H$ $\s$-models,
we split the index $a=(i,\a)$, where the indices $i$ and $\a$ belong
to the subgroup $H\in G$ and the coset $G/H$, respectively.
In \eqn{dualmods} we consider, instead of $E$, a matrix $E_0$
with coset indices only, a restriction requiring that $E_0$ is $G$-invariant.
For coset models the group acts with isotropy and one has to
gauge fix $\dim(H)$ variables among the Lagrange multipliers $v_a$.
Denoting the remaining variables by $x_\a$, we define the $\dim(G/H)$-dimensional square matrices
$N_\pm$ associated with the orthonormal frames, from the relations
\be
L^\a_+ = (M^{-1})^{b\a} \partial_+ v_b =  N_+^{\a\b}  \partial_+ x_\b\ ,
\qquad
L^\a_- =  -(M^{-1})^{\a b}  \partial_- v_b = N_-^{\a\b}  \partial_- x_\b\ .
\label{djkh1}
\ee
The Lorentz transformation that relates them is given by (we write $E_0 = \k_0^T \k_0$)
\be
\L = \k_0 N_+ N_-^{-1}\k_0^{-1} \ .
\label{lool}
\ee

\noindent
As an example consider within the type-IIB supergravity the $AdS_3\times S^3\times T^4$
geometry arising in the near horizon of the D1-D5 brane system. This is supported by an $F_3$ flux
given by the sum of the volume forms of the two group spaces.
The presence of $S^3$ indicates an $SO(4)\simeq SU(2)_L \times SU(2)_R$ isometry
group. Hence we may T-dualize with respect to the full $SO(4)$ or with respect to $SU(2)_L$.
Consider the latter situation first.
One obtains that the fields of the NS sector of dual model are
\be
&& ds^2 = ds^2({\rm AdS_3})  + dr^2  + {r^2\ov 1+ r^2} d\Om_2^2 + ds^2(T^4)\ ,
\nonumber\\
&& B= {r^3\ov 1+r^2} d{\rm Vol}(S^2)\ ,\qq \Phi  = -\ha\ln(1+r^2)\ .
\label{nsnsn}
\ee
The background corresponds to a smooth space, due to the fact that the isometry acts with no isotropy.
In this case the Lorentz transformation is given by (below $ r^2 = x_i x_i$)
\be
\L_{ij} = {r^2-1\ov r^2+1}\ \d_{ij}
- {2\ov r^2+1} (x_i x_j +\e_{ijk}x_k) \quad \Longrightarrow \quad
 \Om = \G_{11} \left(\G_{123} + {\bf x\cdot \G}\ov \sqrt{1+r^2}\right) \ ,
\ee
where $1,2$ and $3$ refer to the directions along the non-Abelian T-dual of $S^3$.
Using \eqn{ppom} we obtain the fluxes
\be
F_0 = 1\ ,\quad F_2= {r^3\ov 1+r^2}\ d{\rm Vol}(S^2)\ ,\quad
F_4 =  - r dr\wedge d{\rm Vol}(AdS_3) + d{\rm Vol}(T^4)\ .
\ee
This is a solution of massive IIA supergravity and
has the residual $SU(2)_R\subset SO(4)$ symmetry.

\noindent
In order to dualize with respect to the full $SO(4)$ symmetry we follow the procedure
outlined above since this group acts with isotropy on the Lagrange multipliers.
We compute the two orthonormal frames corresponding to \eqn{djkh1} and the associated Lorentz
transformation using \eqn{lool} (with $\k_0=\mathbb{I}$). We find that
\be
N_+ ={1\ov x_1 x_3} \left(
       \begin{array}{ccc}
         0 & x_2 & x_3 \\
         0 & x_2^2-x_1^2 & x_2 x_3 \\
         x_1 x_3 & x_2 x_3 & x_3^2 \\
       \end{array}
     \right)\ ,
\quad
\L =\diag(1,-1,-1)\ ,\quad \Om = - \G_2 \G_3\ .
\ee
Hence, $\Om$, is as if we had two successive Abelian T-dualities.
The NS two-form vanishes and the dilaton is computed to be
$\Phi =- \ln (x_1 x_3)$.
Then we compute the Ramond fluxes as
\be
F_1 =2( x_2 dx_3 + x_3 dx_3)\ ,\qquad F_5 = (1+ \star)(F_1 \wedge d{\rm Vol}(T^4))\ .
\label{foo3d}
\ee
This T-dual background is a solution of type-IIB supergravity. The singularity for
$x_1 x_3 =0 $ is associated with the fact that the group acts with isotropy.

\noindent
Finally note that, a manifestly T-duality invariant path integral formulation
of non-Abelian T-duality is known in the broader context of Poisson-Lie T-duality
by doubling the coordinates in a first order $\s$-model action
\cite{Klimcik:1995dy}. It is interesting to embed this in string theory
including the RR-sector fields.

\subsection*{Acknowledgements}

I thank the organizers
for their hospitality as well as the participants for their contribution to the scientific atmosphere of the school.
In addition, I would like to  thank A.P. Polychronakos, D.C. Thompson, E. 'O Colgain and Y. Lozano for
the enjoyable and fruitful collaboration.

 \end{document}